\documentclass[twocolumn,showpacs,preprintnumbers]{revtex4}

\usepackage{graphicx}
\usepackage{dcolumn}
\usepackage{bm}

\begin{document}

\title{Single-cycle megawatt terahertz pulse generation from a
wavelength-scale plasma oscillator driven by ultrashort laser pulses}
\author{Hui-Chun Wu, Zheng-Ming Sheng\footnote{
e-mail: zmsheng@aphy.iphy.ac.cn}, and Jie Zhang}
\affiliation{Laboratory of Optical Physics, Institute of Physics, Chinese Academy of
Science, Beijing 100080, China}
\date{\today }

\begin{abstract}
The tremendous applications of terahertz (THz) spectroscopy and imaging
require THz sources in different parameters. We propose a novel scheme to
generate single-cycle powerful THz pulses by ultrashort intense laser pulses
incident obliquely on a tenuous plasma slab of few THz wavelengths in
thickness. This is made possible by driving a large amplitude electron
plasma wave in the plasma slab, thus producing a net transient current at
the plasma surfaces. Theory and simulations show that such a THz source is
capable of providing megawatt power and field strength of MV/cm, which may
open up new horizons for nonlinear THz science and applications.
\end{abstract}

 \maketitle

\pagebreak

Terahertz spectroscopy can probe the spectral properties of molecules in a
previously inaccessible electromagnetic spectrum \cite{Zhang}. Its
applications include the characterization of semiconductors \cite%
{Grischkowsky} and high-temperature superconductors \cite{Kaindl}, T-ray
imaging of biomedical tissues \cite{Loffler}, cellular structures \cite%
{Mitrofanov} and dielectric substances \cite{Zhang1}, and the manipulation
of bound atoms \cite{Cole}. Most applications are based on the techniques of
THz time-domain spectroscopy (THz-TDS) \cite{Zhang}, where typically
coherent broadband THz pulses in the 2-5 THz frequency bandwidth are
employed. Most broadband pulsed THz sources are based on the excitation of
different materials with ultrashort laser pulses \cite{Zhang}, such as
through photoconduction or optical rectification, but the output power is
limited by the damage threshold of the optical materials used.

So far, there have been no development in the nonlinear regime due to the
lack of sufficiently high power THz sources \cite{Zhang}. To address this
problem a few efforts have been made using ultrashort electron bunches
produced by lasers or accelerators through the mechanisms of transition
radiation and/or synchrotron radiation \cite{Carr}.

Recently, by four-wave mixing in air with a laser intensity of $\sim $10$%
^{14}$ W/cm$^{2}$, single-cycle THz pulses with a field strength greater
than kV/cm can be produced \cite{Cook}, after the laser field-ionization and
plasma generation threshold is surpassed. Whether this kind of THz emission
will saturate for light intensities $>$10$^{15}$ W/cm$^{2}$ is an open
question.

For light intensities $\gg$10$^{14}$ W/cm$^{2}$, the leading part of the
laser pulse can completely ionize the material and the interaction process
becomes a pure laser-plasma interaction. Plasma has no thermal damage
threshold and can sustain extremely intense light. The peak intensity of
lasers today can be as high as 10$^{20}$ W/cm$^{2}$ and in the future is
expected to reach 10$^{23}$ W/cm$^{2}$ \cite{Mourou}. Exploring new THz
emission mechanisms in the context of intense laser plasma interactions may
produce higher power THz sources. It has been shown that laser wakefields
(electron plasma waves driven by the ponderomotive force of a laser pulse)
in inhomogeneous plasmas can radiate high-efficiency THz waves at high power
through linear mode conversion \cite{Sheng,Sheng1}. The THz pulses produced
in this way are generally in multi-cycles and have a negative frequency
chirp.

The present work introduces a new THz emission mechanism in laser plasma
interaction, which can directly generate a single-cycle THz pulse with a
field strength comparable with our earlier proposal \cite{Sheng,Sheng1}. In
addition to the above applications, the single-cycle pulse has special
implications for THz propagation physics and seismic surveys \cite{Dorney}.
In the one-dimensional (1D) case, an electron plasma wave \cite{esarey} in a
uniform plasma is described by $\delta n=\delta n_{p}\exp [i(k_{p}x-\omega
_{p}t)]$, where $\delta n_{p}$ is the density perturbation amplitude, $%
\omega _{p}=\sqrt{ne^{2}/m\varepsilon _{0}}$ is the plasma frequency of the
background plasma of density $n$, $-e$ and $m$ are the electron charge and
mass respectively, and $k_{p}=\omega _{p}/c=2\pi /\lambda _{p}$, where $%
\lambda _{p}$ is the plasma wavelength. This infinite plasma wave can never
emit electromagnetic waves at frequency $\omega _{p}$, since its
displacement current ($\varepsilon _{0}\partial E/\partial t$) exactly
compensates the plasma current ($-env$). However, for a finite plasma wave
of length $L\sim \lambda _{p}$, there could be some electromagnetic
radiation. Firstly, for such a few-wavelength plasma oscillator, its
displacement and plasma currents cannot completely counteract each other, in
particular near the plasma boundaries. Secondly, since the plasma skin depth
of the radiation at frequency $\omega _{p}$ is $k_{p}^{-1}$, which is
comparable to the plasma length $L$, the radiation can tunnel through the
plasma into vacuum. Figure 1(a) is a schematic of this THz wave emission
mechanism.

In the following we give a theoretical analysis of this THz emission
mechanism. For the interaction geometry shown in Fig. 1(a), by Lorentz
transformations we transform all the physical quantities from the laboratory
frame to a moving frame of velocity $c\sin \theta \mathbf{e}_{y}$, where $%
\mathbf{e}_{y}$ is the unit vector along the $y$ direction. An
electromagnetic wave with $\omega ^{\text{L}}=\omega ^{\prime }$, $\mathbf{k}%
^{\text{L}}=(\pm k^{\prime }\cos \theta ,k^{\prime }\sin \theta ,0)$ in the
laboratory frame becomes $\omega ^{\text{M}}=\omega ^{\prime }\cos \theta $,
$\mathbf{k}^{\text{M}}=(\pm k^{\prime }\cos \theta ,0,0)$ in the moving
frame, wherein all the electromagnetic waves propagate along the $\pm x$
directions, i.e. THz emission in the laboratory frame must be in the
specular reflection and laser transmission directions. Plasma (electrons and
ions) in the moving frame streams along $-\mathbf{e}_{y}$, with a
relativistic factor $\gamma ^{\text{M}}=1/\cos \theta $. Following Ref. \cite%
{Lichters} and using the quasi-static approximation \cite{Sprangle}, we
obtain the coupled equations (in SI units)
\begin{equation}
\left( \frac{\partial ^{2}}{\partial x^{2}}-\frac{1}{c^{2}}\frac{\partial
^{2}}{\partial t^{2}}\right) \mathbf{a}_{T}=\frac{\omega _{p}^{2}}{c^{2}}%
\mathbf{s}(x,t),
\end{equation}%
\begin{equation}
\frac{\partial ^{2}\phi }{\partial x^{2}}=\frac{\omega _{p}^{2}}{c^{2}}%
\delta n,
\end{equation}%
\begin{equation}
\delta n=\frac{1}{2\cos \theta }\left[ \frac{1+(\mathbf{a}_{L}-\mathbf{e}%
_{y}\tan \theta )^{2}}{(1/\cos \theta +\phi )}-1\right] ,
\end{equation}%
where $\mathbf{s}(x,t)=-\delta n\tan \theta \mathbf{e}_{y}/\gamma $ is the
THz radiation source, $\mathbf{a}_{T}$ and $\mathbf{a}_{L}$ are the
respective vector potentials of the THz wave and incident laser normalized
by $mc/e$, $\phi $ is the scalar potential of the driven plasma wave
normalized by $mc^{2}/e$, $\delta n$ is the density perturbation of the
plasma wave normalized by the initial plasma density $n$, $\gamma =\sqrt{1+(%
\mathbf{a}_{L}-\tan \theta \mathbf{e}_{y})^{2}+p_{x}^{2}}$ is the electron
relativistic factor, and $p_{x}$ is the electron longitudinal momentum
normalized by $mc$.

The generation of the THz radiation is determined by Eq. (1), from which the
electric field in the laboratory frame is found to be\cite{Lichters}
\begin{equation}
\mathbf{e}_{T}^{\text{L}}(x,t)=\frac{\omega _{p}}{2\omega \cos \theta }%
\int\nolimits_{0}^{L}\frac{dx^{\prime }}{k_{p}^{-1}}\mathbf{s}(x^{\prime
},t-|x-x^{\prime }|/c),
\end{equation}%
where $L$ is the plasma length, $\omega $ the laser frequency, and
normalization with respect to $m\omega c/e$ has been performed. Equation (4)
shows that the THz emission is always $p$-polarized. In the weakly
relativistic case $\phi <<1$, Eqs. (2) and (3) lead to $\delta n\propto
a_{L}^{2}\cos \theta $. In the THz emission process, we assume $\gamma
\simeq \gamma ^{\text{M}}$. Substituting these into Eq. (4) we obtain
\begin{equation}
e_{T}^{\text{L}}\propto n^{1/2}a_{L}^{2}\sin \theta
\end{equation}%
for $L\sim \lambda _{p}$. From Eq. (5) we can see that there is no THz
emission for normal incidence ($\theta =0$), since there is no transverse
electric field component.

To test our proposal and the scaling rule in Eq. (5) we conduct a series of
1D particle-in-cell (PIC) simulations. Taking into account the oblique
incidence of the laser beam, our 1D-PIC code adopts a moving frame as
discussed above \cite{Lichters} and outputs all physics quantities in the
laboratory frame. The initial plasma density is taken to be $n=0.0001n_{c}$,
where $n_{c}=m\varepsilon _{0}\omega ^{2}/e^{2}=1.1\times 10^{21}(\mu $m$%
/\lambda )^{2}$ cm$^{-3}$ is the critical density for the laser pulse of
wavelength $\lambda $ in vacuum. For $\lambda =1$ $\mu $m, $n=1.1\times
10^{17}$ cm$^{-3}$. The corresponding plasma frequency is $\omega _{p}/2\pi
=2.98$ THz, which represents the central frequency of the THz emission. The
plasma wavelength is $\lambda _{p}=\sqrt{n_{c}/n}\lambda =100\lambda $. The
incident laser pulse has a sine-square profile $a_{L}=eE_{L}/m\omega
c=a_{0}\sin ^{2}[\pi (x-ct)/d_{L}]$ for $0\leq x-ct\leq d_{L}$, where $d_{L}$
is the laser pulse duration. Here $a_{0}$ is related to the peak laser
intensity through $I=a_{0}^{2}\cdot 1.37\times 10^{18}(\mu $m$/\lambda )^{2}$
W/cm$^{2}$. The relativistic intensity threshold is reached at $a_{0}=1$.
The laser pulse enters the left boundary of the simulation box with $s$
polarization in order to distinguish it easily from the $p$-polarized THz
emission from the wakefield. For the sine-square laser pulse, the excited
wakefield amplitude is maximum when $d_{L}\approx \lambda _{p}$ \cite{Sheng1}%
. The simulation results also confirm that the THz emission is strongest for
$d_{L}=\lambda _{p}$, thus we always set $d_{L}=\lambda _{p}$ in the
following.

Defining fields $F_{\pm }=(E_{y}\pm cB_{z})/2$ in the moving frame, we see
that $F_{+}$ and $F_{-}$ represent the forward and backward $p$-polarized
electromagnetic waves, respectively. Tracing $F_{+}$ and $F_{-}$ at the
right and left boundaries of the simulation box, we can obtain the temporal
profile of the radiated THz pulses in the reflection and transmission
directions. Figure 1(b) shows the peak field strengths $|F_{\pm }|_{\max }$
of the THz pulses as a function of the plasma length $L$. The laser pulse
parameters are $a_{0}=0.5$, $d_{L}=100\lambda $ and $\theta =45^{\text{o}}$.
The incident laser intensity is about $3.4\times 10^{17}$ W/cm$^{2}$ ($%
\lambda =1$ $\mu $m). We find that the plasma length $L$ should be within $%
[0.25\lambda _{p},2.7\lambda _{p}]$ for intense THz pulses to be generated.
When $L\geq 3\lambda _{p}$, the THz pulse amplitude decreases dramatically.

We also find that the radiated THz pulse is single-cycle. Figures 1(c) and
1(d) show the time evolution of field components $E_{x}$ and $B_{z}$ in the $%
x$ space for $L=100\lambda $. The field $E_{x}$ in Fig. 1(c) includes the
longitudinal field of the wakefield and the electric field of the $p$%
-polarized THz emission. The wakefield is completely localized in the plasma
region, while the electric field of the THz wave is mainly outside the
plasma slab. In Fig. 1(d) $B_{z}$ is the pure magnetic field of the THz
wave. It is obvious that two single-cycle THz pulses are radiated from the
plasma region. Due to the propagation delay of the laser pulse, the pulse in
the backward (reflection) direction is generated earlier than that in the
forward (transmission) direction. For the specific laser wavelength $\lambda
=1$ $\mu $m in Fig. 1(d), the field strength is found to be above $10$
MV/cm. When $L\geq 3\lambda _{p}$, the THz pulse is no longer single-cycle.

Figures 2(a) and 2(b) illustrate the temporal profiles of the THz pulses
shown in Figs. 1(c) and 2(d) together with two other incident angles of $30^{%
\text{o}}$ and $60^{\text{o}}$. The shape of the transmitted THz wave $F_{+}$
is the same as that of the reflected $F_{-}$. For $\theta =30^{\text{o}}$,
the THz pulses have two cycles. Single-cycle THz emission is produced when $%
\theta \geq 45^{\text{o}}$. With increasing incident angle, the number of
cycles included in the THz pulse decreases. Figure 2(c) displays the power
spectra of the THz pulses $F_{-}$. The central frequency is at $3$ THz ($%
\lambda =1$ $\mu $m). The spectrum width increases with the incident angle,
because of the shorter THz duration for the larger $\theta $. The bandwidths
approach 3-6 THz, meeting the requirements for the THz-TDS system. Figure
2(d) shows that the peak field strengths $|F_{\pm }|_{\max }$ are
proportional to $\sin \theta $, which agrees with Eq. (5). There is no THz
emissions for $\theta =0$.

Figures 3(a) and 3(b) show that the THz field strength is
proportional to both the laser intensity, i.e. $a_{L}^{2}$, and the
square root of the plasma density $n$. At lower intensities of
10$^{14}$-10$^{15}$ W/cm$^{2}$, the THz field strength is several
tens of kV/cm, comparable to that generated through four-wave mixing
in air \cite{Cook} under the same intensities.

The above 1D theoretical analysis and 1D-PIC simulations are valid as long
as the laser spot size is large compared with the plasma wavelength. For a
laser beam with a Gaussian profile $\exp (-r^{2}/w_{L}^{2})$ in transverse
space, the 1D model applies for $w_{L}\gg \lambda _{p}$. It is easily
understood that, for $w_{L}<\lambda _{p}$, the radiation source size is
smaller than the radiated wavelength, so that the generated THz wave will
diffract dramatically. In order to have collimated THz emission, $w_{L}\gg
\lambda _{p} $ should be maintained.

To illustrate the multi-dimensional properties of the THz emission, Fig. 4
shows the THz emission in a 2D-PIC simulation. It is found that there are
indeed two single-cycle THz pulses in the reflection and transmission
directions. For $\lambda =1$ $\mu $m, the THz field strength is $42$ MV/cm,
i.e. an intensity of $2.5\times 10^{12}$ W/cm$^{2}$. Since the radiation
radius is about $30\mu $m, this is equivalent to a peak power of $70$ MW.

A uniform plasma slab several tens of microns long can be readily formed
from the gas jet targets commonly used in high-harmonic generation
experiments \cite{Spielmann}. Meanwhile, our numerical simulations show that
THz pulses emitted from a nonuniform plasma with a trapezoid density profile
are similar to those from a uniform plasma slab, provided that the ascending
and decending parts of the trapezoid are also at the THz wavelength scale.
This suggests that the density homogeneity of the plasma slab is not
necessary.

To conclude, we have presented a method for producing single-cycle high
power THz radiation from wavelength-scale plasma oscillators ($L \sim
\lambda_p$). It is emitted by the transient net currents induced at the
plasma surfaces while building-up the plasma oscillators. This mechanism
together with that for THz emission by linear mode conversion in
inhomogeneous plasmas ($L\gg \lambda_p$) \cite{Sheng,Sheng1} provide a
complete picture for interpreting the early experimental observation of THz
emission in intense laser plasma interaction \cite{Hamster}.

\bigskip

\begin{acknowledgments}
Authors are grateful to Prof. Ling-An Wu for careful reading of the
manuscript and improving the text, and Dr. Yu Cang for useful discussions.
This work was supported by the China NNSF (Grants No. 10335020, 10425416,
10390160, 10476033, and 10576035), the National High-Tech ICF Committee in
China, and the Knowledge Innovation Program, CAS.
\end{acknowledgments}

\pagebreak

\pagebreak

\centerline{\Large Figure Captions} \bigskip

FIG. 1 (color online). (a) Schematic of THz emission from a
few-wavelength long plasma oscillator excited by a laser pulse
obliquely incident on a uniform plasma slab. The plasma length is
$L\sim \lambda _{p}$ and $\theta $ is the laser incident angle. THz
waves are radiated in the specular reflection and laser transmission
directions, similar to the optical rectification scheme. (b) The
peak field strengths $|F_{\pm }|_{\max }$ of the radiated THz pulses
as
a function of the plasma length $L$. The plasma is of density $n=0.0001n_{c}$%
. The laser pulse parameters are $a_{0}=0.5$, $d_{L}=100\lambda $ and $%
\theta =45^{\text{o}}$. Spatial-temporal plots of (c) the electric field $%
E_{x}$ and (d) the pure THz magnetic field $B_{z}$ for the plasma
length $L=100\lambda $ (located in $100\lambda \leq x\leq 200\lambda
$). The vertical dashed lines represent the plasma boundaries. The
dashed arrow line marks the laser propagation trajectory. Other
parameters are the same as in (b).

\bigskip

FIG. 2 (color online). Temporal profile, frequency spectrum and
field amplitude of the THz emission. (a) Temporal profiles of the
THz waves $F_{+}$ for incident angles of $\theta =30^{\text{o}}$,
$45^{\text{o}}$ and $60^{\text{o}}$. (b) Temporal profiles of the
THz waves $F_{-}$. Other parameters are the same as in Fig. 1(c,d).
(c) The power spectra of $F_{-}$ in (b). (d) The peak field
strengths $|F_{\pm }|_{\max }$ of the THz pulses as a function of
sin$\theta
$ with $\theta \in \lbrack 0^{\text{o}},15^{\text{o}},30^{\text{o}},45^{%
\text{o}},60^{\text{o}},75^{\text{o}}]$.

\bigskip

FIG. 3 (color online). The scaling rule of the THz emission. (a) The peak field strengths $%
|F_{\pm }|_{\max }$ of the THz pulses as a function of the laser
intensity. The plasma parameters are $n=0.0001n_{c}$, $L=50\lambda $
and the laser
pulse parameters $d_{L}=100\lambda $ and $\theta =45^{\text{o}}$. (b) $%
|F_{\pm }|_{\max }$ of the THz pulses as a function of the electron density $%
n$. The laser pulse parameters are $a_{0}=0.5$ and $\theta =45^{\text{o}}$.
For a given plasma density $n$, we take $L=d_{L}=\lambda _{p}$. The dashed
lines are fitted curves.

\bigskip

FIG. 4 (color online). 2D spatial plot of the pure THz magnetic
field $B_{z}$\ from 2D-PIC simulation. We take $n=0.0025n_{c}$,
corresponding to $\lambda
_{p}=20\lambda $ and $\omega _{p}/2\pi =14.9$ THz. The plasma length is $%
L=25\lambda $. The laser pulse is $s$-polarized, focused on the plasma slab
surface, and has parameters $a_{0}=0.5$, $w_{L}=30\lambda $, $%
d_{L}=20\lambda $ and $\theta =50^{\text{o}}$. The dashed rectangle shows
the plasma region, the solid arrow marks the laser propagation axis, and the
dashed arrows the THz emission directions, one along the laser propagation
and another along the specular reflection direction.

\bigskip

\end{document}